\documentclass[
superscriptaddress,
twocolumn,
amsmath,amssymb,
aps,
prx,
floatfix,
]{revtex4-2}

\usepackage{graphicx}
\usepackage{bm}
\usepackage{hyperref}
\usepackage{upgreek}
\usepackage{float}
\usepackage{enumitem}
\usepackage{braket}
\usepackage{xcolor}
\usepackage{placeins}
\usepackage{tikz}
\usepackage{svg}
\usepackage{mhchem}
\usepackage{siunitx}
\usepackage{romannum}
\usetikzlibrary{quantikz}
\usepackage{multirow}
\setlist{noitemsep}
\usepackage{threeparttable}
\usepackage{makecell}
\usepackage{tabularx}
\newcolumntype{Y}{>{\centering\arraybackslash}X}
\usepackage[T1]{fontenc}

\renewcommand{\u}{\uparrow}
\renewcommand{\d}{\downarrow}

\newcommand{\U}{\Uparrow}
\newcommand{\D}{\Downarrow}

\renewcommand{\P}{${ }^{31}$P}

\begin{document}

\title{Grover's algorithm in a four-qubit silicon processor above the fault-tolerant threshold}

\author{I. Thorvaldson}
\thanks{These authors contributed equally to the work}
\affiliation{Silicon Quantum Computing Pty Ltd., UNSW Sydney, Australia}
\affiliation{Centre of Excellence for Quantum Computation and Communication Technology, UNSW Sydney, Australia}
\author{D. Poulos}
\thanks{These authors contributed equally to the work}
\affiliation{Silicon Quantum Computing Pty Ltd., UNSW Sydney, Australia}
\author{C. M. Moehle}
\thanks{These authors contributed equally to the work}
\affiliation{Silicon Quantum Computing Pty Ltd., UNSW Sydney, Australia}
\author{S. H. Misha}
\affiliation{Silicon Quantum Computing Pty Ltd., UNSW Sydney, Australia}
\author{H. Edlbauer}
\affiliation{Silicon Quantum Computing Pty Ltd., UNSW Sydney, Australia}
\author{J.~Reiner}
\affiliation{Silicon Quantum Computing Pty Ltd., UNSW Sydney, Australia}
\author{H.~Geng}
\affiliation{Silicon Quantum Computing Pty Ltd., UNSW Sydney, Australia}
\author{B. Voisin}
\affiliation{Silicon Quantum Computing Pty Ltd., UNSW Sydney, Australia}
\author{M. T. Jones}
\affiliation{Silicon Quantum Computing Pty Ltd., UNSW Sydney, Australia}
\author{M. B. Donnelly}
\affiliation{Silicon Quantum Computing Pty Ltd., UNSW Sydney, Australia}
\affiliation{Centre of Excellence for Quantum Computation and Communication Technology, UNSW Sydney, Australia}
\author{L. F. Peña}
\affiliation{Silicon Quantum Computing Pty Ltd., UNSW Sydney, Australia}
\author{C. D. Hill}
\affiliation{Silicon Quantum Computing Pty Ltd., UNSW Sydney, Australia}
\affiliation{Centre of Excellence for Quantum Computation and Communication Technology, UNSW Sydney, Australia}
\author{C.~R.~Myers}
\affiliation{Silicon Quantum Computing Pty Ltd., UNSW Sydney, Australia}
\author{J. G. Keizer}
\affiliation{Silicon Quantum Computing Pty Ltd., UNSW Sydney, Australia}
\affiliation{Centre of Excellence for Quantum Computation and Communication Technology, UNSW Sydney, Australia}
\author{Y. Chung}
\affiliation{Silicon Quantum Computing Pty Ltd., UNSW Sydney, Australia}
\author{S. K. Gorman}
\altaffiliation[]{These authors contributed equally to the measurement\\supervision}
\affiliation{Silicon Quantum Computing Pty Ltd., UNSW Sydney, Australia}
\affiliation{Centre of Excellence for Quantum Computation and Communication Technology, UNSW Sydney, Australia}
\author{L. Kranz}
\altaffiliation[]{These authors contributed equally to the measurement\\supervision}
\affiliation{Silicon Quantum Computing Pty Ltd., UNSW Sydney, Australia}
\affiliation{Centre of Excellence for Quantum Computation and Communication Technology, UNSW Sydney, Australia}
\author{M. Y. Simmons}
\email{michelle.simmons@unsw.edu.au}
\affiliation{Silicon Quantum Computing Pty Ltd., UNSW Sydney, Australia}
\affiliation{Centre of Excellence for Quantum Computation and Communication Technology, UNSW Sydney, Australia}

\date{\today}

\begin{abstract}

Spin qubits in silicon are strong contenders for the realization of a practical quantum computer, having demonstrated single and two-qubit gates with fidelities above the fault-tolerant threshold, and entanglement of three qubits. However, maintaining high-fidelity operations while increasing the qubit count remains challenging and therefore only two-qubit algorithms have been executed. Here we utilize a four-qubit silicon processor with all control fidelities above the fault-tolerant threshold and demonstrate a three-qubit Grover’s search algorithm with a $\sim$95\% probability of finding the marked state. Our processor is made of three phosphorus atoms precision-patterned into isotopically pure silicon, which localise one electron. The long coherence times of the qubits enable single-qubit gate fidelities above 99.9\% for all qubits. Moreover, the efficient single-pulse multi-qubit operations enabled by the electron-nuclear hyperfine interaction facilitate controlled-Z gates between all pairs of nuclear spins with fidelities above 99\% when using the electron as an ancilla. These control fidelities, combined with high-fidelity non-demolition readout of all nuclear spins, allow for the creation of a three-qubit Greenberger–Horne–Zeilinger state with 96.2\% fidelity. Looking ahead, coupling neighbouring nuclear spin registers, as the one shown here, via electron-electron exchange may enable larger, fault-tolerant quantum processors.

\end{abstract}

\maketitle

Spin qubits in silicon hold great promise for the realization of large-scale quantum computers due to their long coherence times, compatibility with advanced manufacturing technology and the possibility to operate at elevated (\SI{\sim 1}{K}) temperatures~\cite{Vandersypen_2017,gonzalez2021scaling,Chatterjee_2021,Burkhard_2023}. However, correcting for unavoidable errors requires large numbers of qubits with sufficient quality. The well-known surface code~\cite{Raussendorf_2007, Fowler_2012} demands that the fidelity of every qubit operation within these multi-qubit processors (initialization, readout, and single- and two-qubit control) is above a threshold of approximately 99\%. While high-fidelity initialization, readout, and single- and two-qubit gates have been demonstrated in gate-defined quantum dots~\cite{Takeda_2024,Huang_2024,Yoneda_2018,Yang_2019,Xue_2022,Noiri_2022, Mills_2022,Wu_2024}, combining all of these operations within a single multi-qubit device remains challenging. Reports on the successful implementation of multi-qubit algorithms therefore remain scarce (see Table S1 in Supplementary section \Romannum{1}). First results on two-qubit algorithms~\cite{Watson_2018, Xue_2021} have recently been followed by the implementation of quantum algorithms in two-qubit processors with single- and two-qubit gate fidelities above 99\%~\cite{Xue_2022,Noiri_2022}. Whilst coherent operations have been demonstrated in larger processors (3-6 qubits)~\cite{Takeda_2021,Hendrickx_2021,Philips_2022}, only a three-qubit phase-flip quantum error correction (QEC) code has been executed for devices in which just single-qubit gate fidelities were reported above the fault-tolerant threshold~\cite{Takeda_2022,vanRiggelen_2022}.\\

\begin{figure*}[t!]
  \includegraphics{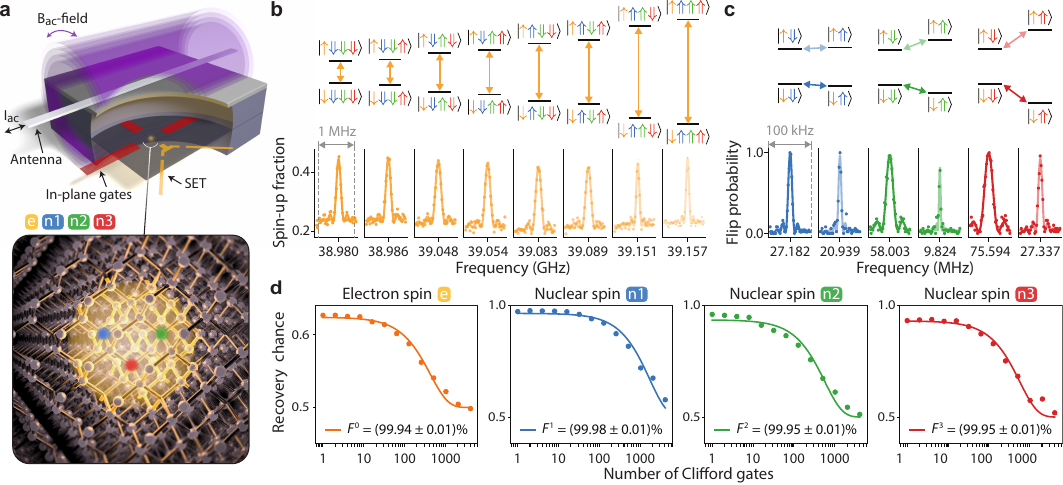}
  \caption{\textbf{Single-qubit operations in the four-qubit processor.} \textbf{a} Schematic illustration of the device consisting of in-plane gates (red), a single electron transistor (SET) charge sensor (yellow) and an antenna (light grey wire) in the top panel. The bottom panel shows an artistic impression of the three phosphorus atoms (blue, green, red) incorporated into the silicon crystal. The attracted electron wavefunction is depicted in yellow. The three nuclear spins and the electron spin define the four-qubit quantum processor. {\bf b} Measured electron spin resonance spectrum displaying 8 resonance frequencies (bottom panel), each corresponding to a different nuclear spin configuration (see energy level diagram in the top panel). {\bf c} Measured nuclear magnetic resonance spectrum (bottom panel) showing all 6 electron spin state-controlled peaks (see energy level diagram in the top panel). {\bf d} Randomized benchmarking decay curves for all four qubits, annotated with the corresponding average physical gate fidelity calculated from the measured Clifford fidelity and the error obtained from the fit. All single-qubit control fidelities surpass 99.9\% fidelity.}
  \label{fig:single_qubit_operations}
\end{figure*}

Phosphorus (\P) atom qubits in silicon (Si) have a number of unique and beneficial properties that can help overcome the challenges of implementing multi-qubit algorithms. The strong natural confinement of atom-based processors allows for the exploitation of the hyperfine interaction between the phosphorus nuclear spins and the bound electron spin. This allows individual qubit addressability~\cite{Hile_2018,Madzik_2022}, while also providing all-to-all qubit connectivity in the form of efficient multi-qubit gates. The latter leverages the fact that a single gate on the electron spin can entangle multiple nuclear spins. This reduces the number of operations needed to execute quantum algorithms. In addition, the nuclear spins have long coherence times~\cite{Muhonen_2014} and can be read out with high fidelity via the process of quantum non-demolition readout~\cite{Pla_2013}. Phosphorus atom qubits in silicon have been realized using ion implantation with high fidelity single and two-qubit gate operations and the recent demonstration of entanglement between two nuclear spins and one electron spin~\cite{Muhonen_2015,Madzik_2022}. Scaling to larger error-corrected architectures requires precision control over the placement of the phosphorus atom qubits, which can be achieved by scanning tunnelling microscopy (STM) lithography~\cite{Fuechsle_2012,Reiner_2024}.\\

Here we demonstrate full coherent control over a precision-manufactured four-qubit processor in Si defined by three phosphorus nuclear spin qubits and one electron spin qubit. We achieve single-qubit gate fidelities for all four individual qubits of ($99.94\pm0.01)\%$, ($99.98\pm0.01)\%$, ($99.95\pm0.01)\%$ and ($99.95\pm0.01)\%$. In addition, we demonstrate two-qubit controlled-Z (CZ) gates between all pairs of nuclear spins with fidelities of ($99.65\pm0.35)\%$, ($99.49\pm0.39)\%$ and ($99.32\pm0.22)\%$, as well as readout of all nuclear spin qubits with a fidelity above 99\% after post-selection. We exploit these high-fidelity ($>$99\%) operations, all achieved within the same device, and the all-to-all qubit connectivity in the processor to produce Bell states and a three-qubit Greenberger-Horne-Zeilinger (GHZ) state with fidelities above 96\%. Finally, we benchmark our four-qubit processor by executing a Grover's search algorithm on the three nuclear spin qubits with a $(94.57\pm2.63)\%$ average success probability of finding the marked state compared with the ideal algorithm fidelity. This constitutes one of the most successful implementations of this algorithm in any qubit platform so far.

\section{Single-qubit operations}

\begin{figure*}[t!]
\centering
\includegraphics{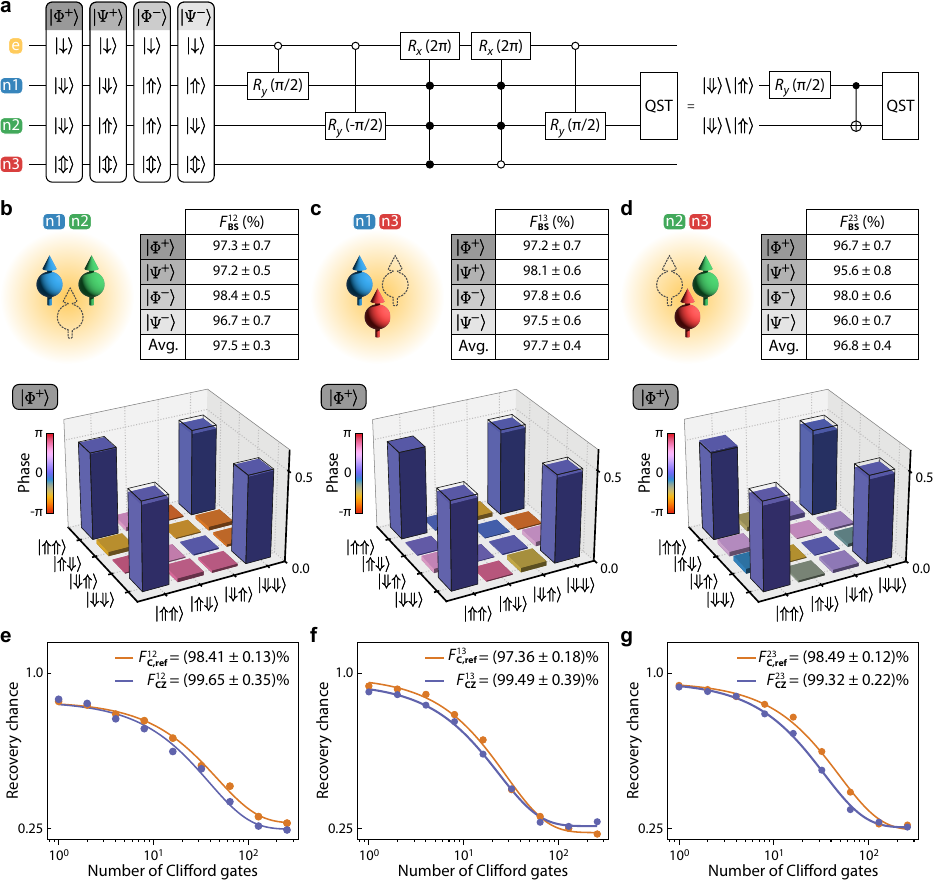}
\caption{\textbf{Bell state tomography and two-qubit randomized benchmarking.} \textbf{a} Circuit diagram used to construct the Bell states for nuclear spin 1 and 2. The table shows the input states corresponding to the final Bell state. \textbf{b-d} Fidelities for all Bell states and reconstructed density matrices for $\ket{\Phi^+}$ obtained from full-basis quantum state tomography, for all pairs of nuclear spins. The errors are obtained from Monte Carlo bootstrap re-sampling and represent $1\sigma$ from the mean. Panel \textbf{b} corresponds to nuclear spin 1 and 2, panel \textbf{c} to nuclear spin 1 and 3, and panel \textbf{d} to nuclear spin 2 and 3 (see schematics). \textbf{e-g} Two-qubit randomized benchmarking (orange) and two-qubit interleaved randomized benchmarking (purple), for all pairs of nuclear spins, corresponding to the same pairs of spins studied in b-d. The physical CZ gate fidelites ($F_{\mathrm{CZ}}^{12}$, $F_{\mathrm{CZ}}^{13}$ and $F_{\mathrm{CZ}}^{23}$) are calculated from the non-interleaved reference Clifford fidelities ($F_{\mathrm{C,ref}}^{12}$, $F_{\mathrm{C,ref}}^{13}$ and $F_{\mathrm{C,ref}}^{23}$) and the interleaved Clifford fidelities (obtained from the fits to the purple data points). The errors are obtained from the fits. The physical CZ gate fidelities for all pairs of nuclear spins are above the fault-tolerant threshold.}
\label{fig:two_qubit_operations}
\end{figure*}

The multi-qubit processor studied in this work is formed by patterning three \P~atoms into isotopically purified \ce{^{28}Si} with atomic precision using STM hydrogen lithography, as previously described for the same device by Reiner {\it et. al.}~\cite{Reiner_2024}. Highly phosphorus doped silicon in-plane gates allow control of the electrostatic environment of the P atoms (highlighted in red in the schematic illustration of the device in the top panel of Fig.~\ref{fig:single_qubit_operations}a). An electron can be loaded onto the P atoms from a nearby, tunnel-coupled single-electron transistor (SET, yellow) that also serves as a charge sensor. To control the nuclear spins (basis states $\ket{\D}$,$\ket{\U}$) and the electron spin (basis states $\ket{\d}$,$\ket{\u}$), a broadband antenna (grey) is placed on top of the device, delivering the radio-frequency and microwave signals for nuclear magnetic resonance (NMR) and for electron spin resonance (ESR), respectively. An artistic impression of the four-qubit processor is shown in the bottom panel of Fig.~\ref{fig:single_qubit_operations}a, where the electron wave function (yellow) spreads over the three P atoms (blue, green and red) that are embedded in the silicon crystal. In the following we use label 0 for the electron spin qubit and labels 1, 2 and 3 for the nuclear spin qubits (blue, green and red, respectively).\\

Electron spin initialization and readout is performed via a ramped technique~\cite{Keith_2022} at a dilution refrigerator base temperature of \SI{15}{mK}, with an applied magnetic field of \SI{1.45}{T}. Quantum non-demolition readout of the nuclear spins is achieved with fidelities above 99\% after post-selection as shown in Supplementary section \Romannum{2} with nuclear spin initialization shown in Supplementary section \Romannum{3}.\\

When an electron is loaded onto the multi-nuclear spin register, the electron spin interacts with all the nuclear spins through the contact hyperfine interaction, causing the ESR frequency to depend on the state of each of the nuclear spins. Figure~\ref{fig:single_qubit_operations}b (bottom) shows an ESR spectrum with 8 resonance peaks, where each peak corresponds to a different configuration of the nuclear spins (see energy level diagram in the top panel). From the ESR peak separations we find hyperfine interaction strengths of $A_1=\SI{6}{MHz}$, $A_2=\SI{68}{MHz}$ and $A_3=\SI{103}{MHz}$. The presence of the hyperfine interaction also allows each nuclear spin to be addressed separately, with the NMR frequency depending on the targeted nuclear spin and the state of the electron spin. An NMR spectrum displaying all 6 expected peaks is shown in the bottom panel of Fig.~\ref{fig:single_qubit_operations}c (see energy level diagram in the top panel).\\

Having established full individual addressability of the electron spin and the three nuclear spins, we measure the dephasing time of each qubit using a Ramsey experiment. We find $T_2^{*}=\SI{28.1}{\mu s}$ for the electron spin and $\SI{1.26}{ms}$, $\SI{0.49}{ms}$, $\SI{0.60}{ms}$ for nuclear spins 1, 2 and 3, respectively (see Supplementary section \Romannum{4}). To measure the dephasing time of the electron spin, we initialize all nuclear spins into the $\ket{\D\D\D}$ state before applying the ESR pulses conditional on that nuclear spin configuration (for all other nuclear spin configurations we find similar $T_2^{*}$ values, see Supplementary section \Romannum{4}). Based on these dephasing times, Rabi dephasing times and the Rabi frequencies (also shown in Supplementary section \Romannum{4}) we obtain qubit quality factors $Q = T_2^* \cdot f_{\mathrm{Rabi}}$ of 4.82, 14.14, 11.84 and 18.86, as well as gate quality factors of $Q = T_2^{\mathrm{Rabi}} \cdot f_{\mathrm{Rabi}}$ of 34, 797, 84 and 69, for the electron spin, nuclear spin 1, nuclear spin 2 and nuclear spin 3, respectively.\\

Next, we characterize the control fidelity of all single-qubit operations by means of randomized benchmarking (RB). We achieve physical gate fidelities, $F^{i}$, above 99.9\% for all four qubits ($i=0,1,2,3$) as displayed in Fig.~\ref{fig:single_qubit_operations}d. Randomized benchmarking for the electron spin is performed with the nuclear spins initialized into the $\ket{\D\D\D}$ state (all other nuclear spin configurations also yield fidelities above $99.9\%$, see Supplementary section \Romannum{5}), whilst RB for the nuclear spins is performed with the electron spin initialized into the $\ket{\d}$ state. With high-fidelity nuclear spin readout and all single-qubit gate fidelities surpassing the fault tolerant threshold, we proceed to create entanglement between the nuclear spins. 

\section{Two- and three-qubit entanglement}

To entangle two of the nuclear spins, we exploit the hyperfine interaction between the electron spin and each of the nuclear spins. Here, simply by enacting a $2\pi$-rotation of the electron spin conditional on the configuration of the nuclear spins, we implement a geometric CZ gate between the nuclear spins~\cite{Waldherr_2014,Madzik_2022}. To illustrate this, starting with the control nuclear spin in state $\ket{\D}$, the target nuclear spin in $(\ket{\D}+\ket{\U})/\sqrt{2}$ and the third nuclear spin in $\ket{\D}$, an ESR $2\pi$-pulse conditional on $\ket{\D\D\D}$ flips the target nuclear spin by $\SI{180}{\degree}$ around the $z$-axis of its Bloch sphere. If, on the other hand, the control state was in state $\ket{\U}$, the same ESR pulse would not affect the target nuclear spin. To implement this gate irrespective of the state of the third nuclear spin, an additional ESR $2\pi$-pulse conditional on $\ket{\D\D\U}$ can be applied, creating a two-qubit CZ gate. Inserting the two ESR pulses in between a $-\pi/2$ rotation and a $\pi/2$ rotation of the target nuclear spin, results in a nuclear controlled-NOT (CNOT) gate.\\

\begin{figure}[t!]
\centering
\includegraphics[width=1.0\columnwidth]{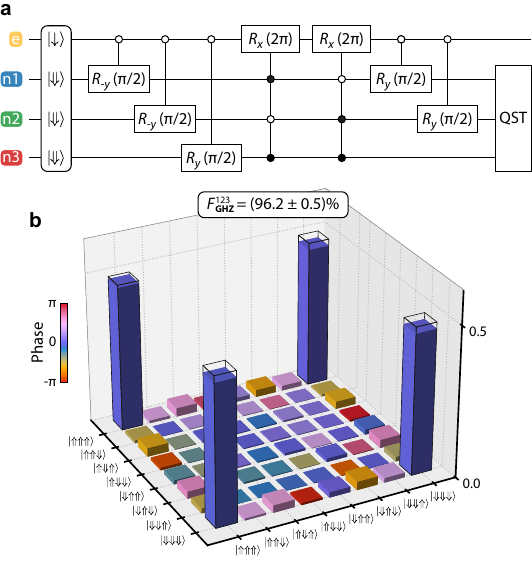}
\caption{\textbf{Three-qubit Greenberger–Horne–Zeilinger state tomography.} \textbf{a} Circuit diagram used to construct the Greenberger–Horne–Zeilinger (GHZ) state. \textbf{b} Reconstructed density matrix of the GHZ state and corresponding state fidelity ($F_{\textrm{GHZ}}^{123} = (96.2 \pm 0.5) \%$) as obtained from full-basis quantum state tomography (QST). The error is obtained from Monte Carlo bootstrap re-sampling and represents $1\sigma$ from the mean.}
\label{fig:GHZ}
\end{figure}

We use this gate to create all of the four Bell states, $\ket{\Phi^{\pm}}=(\ket{\D\D}\pm\ket{\U\U})/\sqrt{2}$, $\ket{\Psi^{\pm}}=(\ket{\D\U}\pm\ket{\U\D})/\sqrt{2}$, for each pair of nuclear spins (see Fig.~\ref{fig:two_qubit_operations}a for the circuit diagram for nuclear spins 1 and 2). At the end of each measurement we perform full-basis quantum state tomography (QST) to reconstruct the density matrix, $\rho^{ij}$, and obtain the corresponding Bell state fidelity from $F_{\mathrm{BS}}^{ij}=\braket{\psi|\rho^{ij}|\psi}$, where $\psi$ is the target Bell state and $i,j=1,2,3$ label the nuclear spins (Supplementary section \Romannum{6}). Figure~\ref{fig:two_qubit_operations}b-d shows the reconstructed density matrices for $\ket{\Phi^{+}}$ for each pair of nuclear spins, with the Bell state fidelities listed in the tables above with state preparation and measurement (SPAM) errors included. The density matrices for $\ket{\Phi^{-}}$ and $\ket{\Psi^{\pm}}$ are shown in the Supplementary section \Romannum{7}. We achieve average Bell state fidelities of $F_{\mathrm{BS}}^{12}=(97.5\pm 0.3)\%$, $F_{\mathrm{BS}}^{13}=(97.7\pm 0.4)\%$, and $F_{\mathrm{BS}}^{23}=(96.8\pm 0.4)\%$ for the three pairs of nuclear spins, among the highest fidelities that have been reported for spin qubits in Si~\cite{Dehollain_2016,Mills_2022}.\\

\begin{figure*}[t!]
\centering
\includegraphics{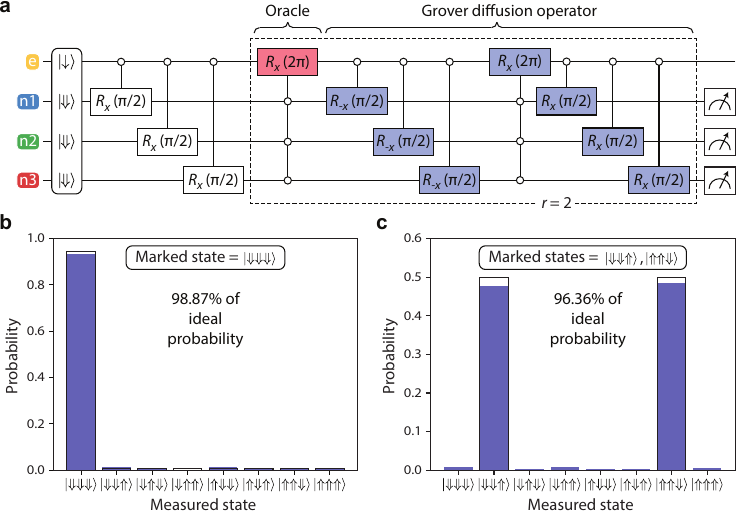}
\caption{\textbf{Three-qubit Grover's algorithm.} \textbf{a} Circuit diagram implementing Grover's algorithm on three nuclear spins. The oracle is highlighted in red (here marking the $\ket{\D\D\D}$ state) and the Grover diffusion operator is highlighted in blue. \textbf{b} Measurement result when using $\ket{\D\D\D}$ as the marked state and performing $r=2$ Grover iterations. \textbf{c} Measurement result when marking the states $\ket{\D\D\U}$ and $\ket{\U\U\D}$, and performing $r=1$ Grover iteration.}
\label{fig:grovers}
\end{figure*}

Bell state fidelities are affected by SPAM errors, single- and two-qubit gate errors, and errors that occur when qubits idle. To independently quantify the fidelity of the CZ gate, we perform two-qubit RB and two-qubit interleaved RB with the CZ gate as the interleaved gate (Supplementary section \Romannum{8}). As shown in Fig.~\ref{fig:two_qubit_operations}e-g, we find CZ gate fidelities of $F_{\mathrm{CZ}}^{12}=(99.65\pm 0.35)\%$, $F_{\mathrm{CZ}}^{13}=(99.49\pm 0.39)\%$ and $F_{\mathrm{CZ}}^{23}=(99.32\pm 0.22)\%$ for the three pairs of nuclear spins. In Supplementary section \Romannum{9} we show that the residual infidelities of the CZ gates can be explained by quasistatic variation in the electron energy splitting, which has been drastically reduced in this device by using isotopically pure silicon-28 material grown at low temperature using molecular beam epitaxy~\cite{Reiner_2024}.
Two-qubit gate fidelities above the fault-tolerant threshold remain scarce in Si spin qubits and have only recently been reported~\cite{Xue_2022, Noiri_2022, Mills_2022, Madzik_2022, Huang_2024, Wu_2024}.\\

As a final demonstration of our ability to create entangled states, we entangle all three nuclear spins to create a GHZ state using the circuit in Fig.~\ref{fig:GHZ}a. The reconstructed density matrix obtained from full-basis QST is shown in Fig.~\ref{fig:GHZ}b. We achieve a fidelity of $F_{\mathrm{GHZ}}^{123}=(96.2\pm 0.5)\%$ (including SPAM errors), the highest GHZ state fidelity reported for semiconductor spin qubits to-date (see Table S1 in Supplementary section \Romannum{1}).\\

\section{Grover's search algorithm}

Finally, we benchmark our 4-qubit quantum processor by executing the well-known Grover's search algorithm~\cite{Grover_1996}, using the corresponding circuit shown in Fig.~\ref{fig:grovers}a. In general, this algorithm finds a specific bit string, $x_m$, in the domain $x$ of a function $f$, where $f$ is defined such that it gives $f(x_m)=1$ and $f(x_i)=0$ for all other $x_i \neq x_m$. In our case, the domain consists of the eight binary values $\{000, 001, \dots, 111\}$, which correspond to the eight possible nuclear spin states $\{\ket{\D\D\D}, \ket{\D\D\U}, \dots, \ket{\U\U\U}\}$. Grover's algorithm works by accessing $f$ with a unitary operator (called oracle), $U_{x_m}$, which performs the action $U_{x_m}\ket{x}=(-1)^{f(x)}\ket{x}$. That is, the searched-for state ($x_m$) is marked with a negative phase, while all other states are left unchanged. Taking advantage of the all-to-all connectivity in our processor, this oracle operation can be performed on the three nuclear spins by applying a single $2\pi$-rotation of the electron spin at the ESR frequency corresponding to $x_m$ (highlighted in red in the circuit diagram in Fig.~\ref{fig:grovers}a). To find the marked state with high probability, the Grover iteration consisting of the oracle and the Grover diffusion operator (highlighted in blue in the circuit diagram and implementing the unitary $2I^{\otimes3}-(\ket{+}\bra{+})^{\otimes 3}$) must be applied multiple times. Note that the Grover diffusion operator also benefits from all-to-all connectivity, only requiring a single entangling gate. For $n=3$ qubits, the optimal number of repetitions is $r=2$, which can be found using $r = \textrm{argmax}_r \sin^2 \big[(2 r + 1)\arcsin\left(2^{-n}\right)\big]$, where argmax takes the earliest local maximum~\cite{Boyer_1998}.\\

In Fig.~\ref{fig:grovers}b we demonstrate the final measurement outcome of Grover's algorithm executed on the three nuclear spins when using $\ket{\D\D\D}$ as the marked state. The algorithm finds this state with a probability of 93.46\%, which corresponds to 98.87\% of the ideal probability (94.53\%) of finding the marked state with $r=2$ Grover iterations. We also run the algorithm for all other marked states (Supplementary section \Romannum{10}) and find an average probability of ($89.40\pm2.49)\%$ of finding the marked state, which corresponds to ($94.57\pm2.63)\%$ of the ideal value. The uncertainties represent $1\sigma$ from the mean taken over all marked states. This average fidelity is in good agreement with the predicted fidelity based on considering all the errors during qubit operations characterized in this work, as shown in Supplementary section \Romannum{11}. In addition, we show that the algorithm can be run with two marked states using $r=1$, which we would expect to achieve a success probability of 100\%. Here, we achieve a probability of $96.36\%$ to find the two marked states (see Fig.~\ref{fig:grovers}c). These results represent one of the most successful implementations of Grover's algorithm among any qubit platform to-date (see Supplementary section \Romannum{12}).\\

\section{Conclusions}
We have shown full coherent control in a four-qubit silicon processor consisting of three nuclear spins and one electron spin. The all-to-all qubit connectivity along with the long coherence times of the spin qubits allowed us to obtain control fidelities above the fault-tolerant threshold and to successfully execute a three-qubit Grover's search algorithm with high accuracy.\\

While in this work we used the electron spin to provide connectivity and to efficiently implement multi-qubit gates, it can also be used to couple neighbouring nuclear spin registers via the electron-electron exchange interaction. Exciting progress has been made in this direction~\cite{He_2019, Madzik_2021, Stemp_2023} and we anticipate the advent of quantum processors consisting of multiple connected registers in the near future. At the same time, as the placement precision of the STM approaches the atomic limit~\cite{Simmons_2019,Ivie_2021,Wyrick_2022}, we anticipate future generations of devices with precisely engineered hyperfine interactions and tunnel couplings.\\

\FloatBarrier

{\bf Acknowledgements: } The research outlined in this paper, as performed by all authors, was conducted and supported by \textit{Silicon Quantum Computing Pty Ltd} [ACN 619 102 608], Australian Research Council Centre of Excellence for Quantum Computation and Communication Technology (project number CE170100012). M.Y.S. acknowledges an Australian Research Council Laureate Fellowship. We acknowledge the FPGA measurement hardware deployment and support from A.~Bridger, V.~Bardell, D.~Antliff, C.~Brown and A.~Sutherland. We acknowledge the measurement software deployment and support from R.~Buckland, S.~Findlay, O.~Gorman and O.~Cowan.\\

{\bf Author contributions: } S.~H.~M., J.~R. and Y.~C. fabricated the device under the supervision of J.~G.~K.; I.~T., D.~P. and C.~M.~M. measured the device under the supervision of S.~K.~G. and L.~K.; I.~T., D.~P., C.~M.~M. and M.~B.~D. analyzed the data; H.~E., H.~G, B.~V., M.~T.~J. and L.~F.~P. contributed to optimizing fabrication, measurements or the experimental setup; C.~D.~H. and C.~R.~M. assisted with the algorithm design; The manuscript was written by C.~M.~M., I.~T., L.~K, S.~K.~G and M.~Y.~S. with input from all authors;  M.~Y.~S. supervised the overall project.\\

{\bf Competing Interests: } M.~Y.~S. is a director of the company Silicon Quantum Computing Pty Ltd. I.~T., D.~P., C.~M.~M., S.~H.~M., H.~E., J.~R., H.~G., B.~V., M.~T.~J., M.~B.~D., L.~F.~P., C.~D.~H., C.~R.~M., J.~G.~K., Y.~C., S.~K.~G., L.~K., and M.~Y.~S. (all authors) declare equity interest in Silicon Quantum Computing Pty Ltd.\\

\section*{Methods}
{\bf Fabrication: } The device was fabricated using hydrogen STM lithography on a silicon chip with a 45 nm layer of isotopically purified silicon-28 ($\approx$200 ppm of residual Si-29 atoms). This epitaxial buffer layer decouples the device from the nuclear spin bath of the natural silicon substrate. After patterning the device, the sample is dosed with phosphine gas, followed by an incorporation anneal at \SI{350}{^\circ C} for 60 seconds to incorporate P atoms into the silicon crystal lattice. A 45 nm epitaxial layer of Si-28 is then grown at \SI{250}{^\circ C} and at a rate of 0.15 nm/min to ensure high-quality epitaxy. The buffer layer separates the qubits from any charged defects on the silicon surface and, together with the encapsulation layer, provide a monolithic qubit environment with minimal levels of spin and charge noise~\cite{Kranz_2020}. An STM image of the device is shown in Figure 1 c,d in~\cite{Reiner_2024}, where the same device as studied in this work was also studied.\\

{\bf Data Availability: } The raw data used in this work, along with code used to analyse this data and produce the figures in this work, is available at \href{http://doi.org/10.5281/zenodo.14214375}{http://doi.org/10.5281/zenodo.14214375}.\\

\pagebreak
\newpage

\renewcommand{\thefigure}{S\arabic{figure}}%
\renewcommand{\thetable}{S\Roman{table}}%

\setcounter{page}{0}
\setcounter{table}{0}
\setcounter{figure}{0}
\setcounter{section}{0}
\setcounter{equation}{0}

\newpage
\onecolumngrid
\section*{Supplementary Information}

\section{State-of-the-art semiconductor spin qubit processors}

\begin{table*}[h!]
\caption{\textbf{Comparison of state-of-the-art semiconductor spin qubit quantum processors using gate-defined quantum dots in Si/SiGe and in Ge/SiGe and using multi-nuclear Si:P spin registers.} We only include processors that have demonstrated an algorithm or QEC (for two-qubit processors we only include those with single and two-qubit gate fidelities above 99\%). Furthermore, we include the processor with the largest number of coherently controlled qubits using gate-defined quantum dots in Si and using Si:P spin registers. ``SPAM'' stands for ``state preparation and measurement'', ``VQE'' for ``variational quantum eigensolver'', and ``DJ'' for ``Deutsch-Jozsa''. \label{tab:comp}}
\footnotesize
\hspace*{-0.6cm}
\begin{threeparttable}[b]
{\renewcommand{\arraystretch}{1.2}
\begin{tabularx}{1.05\textwidth}{| c || Y | Y | Y | Y | Y | Y | Y |}

	\hline

    \makecell{\Gape[10pt]{Reference}} &  
    Xue~\cite{Xue_2022}&
    Noiri~\cite{Noiri_2022}&
    Takeda~\cite{Takeda_2022}&
    Philips~\cite{Philips_2022}&
    \makecell{Hendrickx\\\cite{Hendrickx_2021}\\Van Riggelen\\\cite{vanRiggelen_2022}}&
    Madzik~\cite{Madzik_2022}&
    This work\\
    
    \hline

        \makecell{\vspace{-0.2cm}\\Year\\\vspace{-0.2cm}}&  
        2022&
        2022&
        2022&
        2022&
        2021/2022&
        2022&
        2024\\
    \hline
    
    \makecell{\vspace{-0.2cm}\\Platform\\\vspace{-0.2cm}}&
        Si/SiGe&
        Si/SiGe&
        Si/SiGe&
        Si/SiGe&
        Ge/SiGe&
        Si:P&
        Si:P\\
    \hline

    \makecell{\vspace{-0.2cm}\\Qubits\\\vspace{-0.2cm}}&
        2 (electrons)&
        2 (electrons)&
        3 (electrons)&
        6 (electrons)&
        4 (holes)&
        3 (n-n-e)&
        4 (n-n-n-e)\\
    \hline
    
    \makecell{\vspace{-0.2cm}\\SPAM \\fidelity (\%)\\\vspace{-0.2cm}}&
        -
        &
        74.25\tnote{$a$}
        &
        -
        &
        -
        &
        -
        &
        98.95\tnote{$a$} (n)
        &
        99.42 to 99.57 (n)\\
    \hline

    \makecell{\vspace{-0.2cm}\\Rabi\\ visibility (\%)\\\vspace{-0.2cm}}
        &
        -
        &
        -
        &
        70 to 85\tnote{$b$}
        &
        93.5 to 98\tnote{$c$}
        &
        60 to 75\tnote{$b$}
        &
        -
        &
        92 to 99 (n)
       \\
    \hline

    \makecell{\vspace{-0.2cm}\\Single-qubit\\gate fidelity (\%)\\\vspace{-0.2cm}}&
        {99.71 to 99.74}
        &
        99.84 to 99.84
        &
        99.68 to 99.77
        &
        99.77 to 99.96
        &
        99.40 to 99.88
        &
        99.46 to 99.91 (n)
        &
        99.95 to 99.98 (n)
        \\
    \hline

    \makecell{\vspace{-0.2cm}\\Two-qubit\\gate fidelity (\%)\\\vspace{-0.2cm}}&
        99.65
        &
        99.51
        &
        -
        &
        -
        &
        -
        &
        99.37 (n-n)
        &
        99.32 to 99.65 (n-n)
        \\
    \hline

        \makecell{\vspace{-0.2cm}\\Bell state\\fidelity (\%)\\\vspace{-0.2cm}}&
        \makecell{98.1\tnote{$d$}\\ (w/o SPAM)}
        &
        \makecell{96.5\\ (w/o SPAM)}
        &
        -
        &
        78.0 to 91.3
        &
        -
        &
        93.4 (n-n)
        &
        96.8 to 97.7 (n-n)
        \\
    \hline

    \makecell{\vspace{-0.2cm}\\Three-qubit GHZ\\state fidelity (\%)\\\vspace{-0.2cm}}&
        N/A
        &
        N/A
        &
        \makecell{86.6\\ (w/o SPAM)}
        &
        52.7 to 67.2
        &
        -
        &
        92.5\tnote{$e$} (n-n-e)
        &
        96.2 (n-n-n)
        \\
    \hline
    
    \makecell{\vspace{-0.2cm}\\Demonstration of\\ algorithm or QEC\\\vspace{-0.2cm}}&
        \makecell{Two-qubit\\ VQE\\ algorithm}&
        \makecell{Two-qubit DJ\\and Grover's\\algorithm}&
        \makecell{Three-qubit\\ phase-flip\\ QEC code}&
        -
        &
        \makecell{Three-qubit\\ phase-flip\\ QEC code}
        &
        -
        &
        \makecell{Three-qubit\\ Grover's\\ algorithm}\\
    \hline

	\hline
\end{tabularx}
}

\begin{tablenotes}
\item [$a$] Average two-qubit SPAM fidelity as stated in references~\cite{Noiri_2022,Madzik_2022}.
\item [$b$] Rabi visibility estimated from Extended Data Fig.\ 2b-d in~\cite{Takeda_2022} and Fig.\ 1f in~\cite{Hendrickx_2021}. 
\item [$c$] Values for operating individual qubits; when initializing other qubits in the device the Rabi visibilities decrease as stated in reference~\cite{Philips_2022}.
\item [$d$] From simulation (rather than measurement) as stated in reference~\cite{Xue_2022}.
\item [$e$] From return probability (rather than quantum state tomography) as stated in reference~\cite{Madzik_2022}.

\end{tablenotes}
\end{threeparttable}
\end{table*}

\section{Nuclear spin non-demolition readout~\label{sec:supp_readout}}

\begin{figure*}[t!]
  \includegraphics[width=1.0\textwidth]{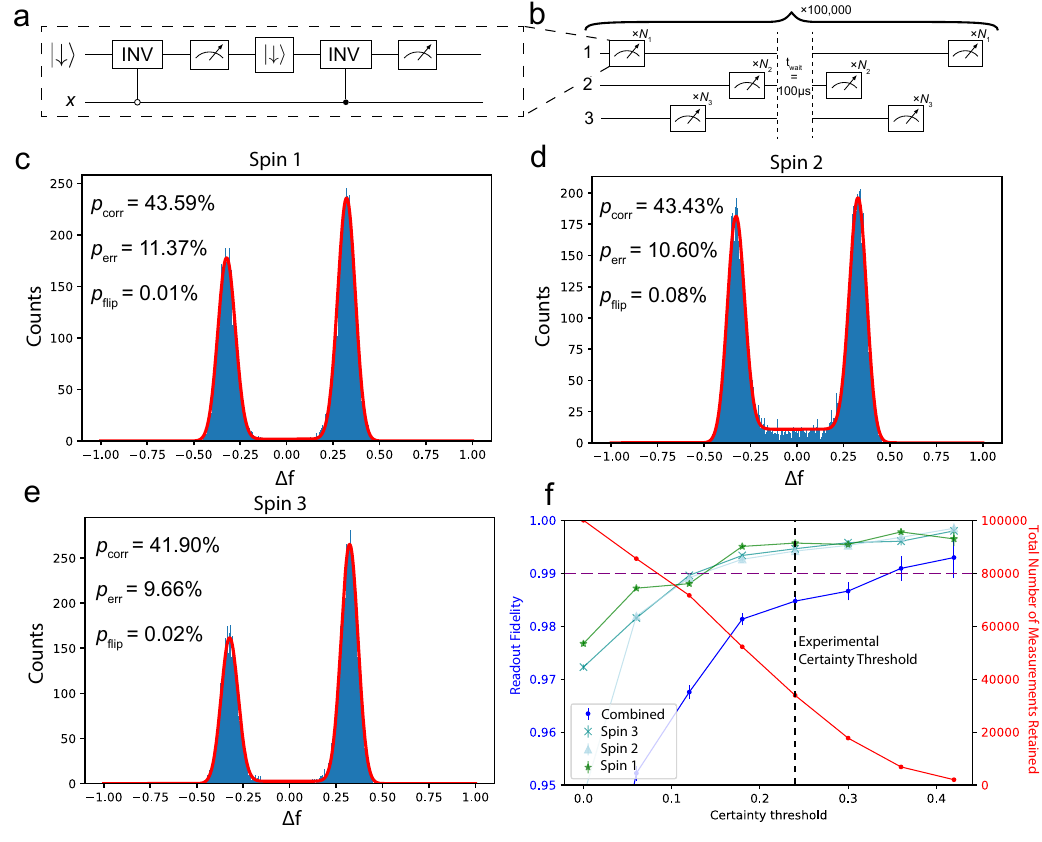}
  \caption{\textbf{Optimization of nuclear spin non-demolition readout.} \textbf{a} Circuit used to perform a single shot of readout of one of the nuclear spins, with $x=1,2,3$ being the label for the nuclear spin that is read out. \textbf{b} Circuit used to measure the readout fidelity of the nuclear spins. This circuit is repeated 100,000 times to obtain the readout fidelities. \textbf{c-e} Histograms of $\Delta f$ for 10,000 repeated readout measurements for all three individual nuclear spins. Each measurement involves taking $N=200$ shots of the corresponding nuclear spin. Included are fits from a Markov chain model (red lines) to extract $p_{\mathrm{corr}}$, $p_{\mathrm{err}}$, and $p_{\mathrm{flip}}$. \textbf{f} Individual and combined nuclear spin readout fidelities (left axis) and number of measurements retained (right axis) as a function of the certainty threshold for ($N_1$,$N_2$,$N_3$)=(24,18,24). Note that different certainty thresholds are used for each nuclear spin, with the threshold displayed on the x-axis being scaled by $(\frac{1}{2}, 1, \frac{2}{3})$ for spins 1, 2 and 3 respectively. The error bars for (f) are calculated according to $\sigma_f = \sqrt{1 + 4 f (1-f) n}\ /2(n+1)$, representing the range of true fidelities for which the observed fidelity lies within $1 \sigma$ of the relevant binomial distribution, as defined in this section.}
  \label{fig:supp_nondemo}
\end{figure*}

The nuclear spins are read out via the electron spin (a detailed description of electron spin readout is provided in ~\cite{Reiner_2024}). This nuclear spin measurement is quantum non-demolition, i.e. the nuclear spin remains in the projected measured state after the measurement operation. For every nuclear spin we perform $N$ readout shots, each shot consisting of the following operations (see Fig.\ \ref{fig:supp_nondemo}a): initialization of the electron spin into the $\ket{\d}$ state, adiabatic inversion of the electron spin conditional on the nuclear spin being in the $\ket{\D}$ state, electron spin readout, initialization of the electron spin into the $\ket{\d}$ state, adiabatic inversion of the electron spin conditional on the nuclear spin being in the $\ket{\U}$ state, and finally electron spin readout. From this sequence of measurements, we obtain the fraction of shots detecting nuclear spin down ($f_{\D}=N_{\D}/N$) and nuclear spin up ($f_{\U}=N_{\U}/N$), where $N_{\D/\U}$ is the number of electron spin up events detected for the specific nuclear spin state. If the value of $\Delta f=f_{\U}-f_{\D}$ is positive we assign a nuclear state $\ket{\U}$, and for $\Delta f \le 0$ we assign a nuclear state $\ket{\D}$. For repeated measurements (each consisting of $N$ readout shots), we can form a histogram of the observed values of $\Delta f$. Examples of this are shown in Fig.\ \ref{fig:supp_nondemo}c-e, where we observe two well separated Gaussian peaks corresponding to the nuclear spin state $\ket{\U}$ and $\ket{\D}$. To maximize the readout fidelity for every nuclear spin for the tomography and Grover's algorithm data obtained in this work, we optimize the number of readout shots and postselect the observed nuclear readouts, keeping only those readouts where $|\Delta f|$ lies above a defined ``certainty threshold'' ($f_{\mathrm{th}}$). This certainty threshold is designed to remove measurements where the nuclear spin flipped during the nuclear non-demolition readout.\\

To find the optimal number of readout shots for each nuclear spin ($N_1$,$N_2$,$N_3$), we model the nuclear spin readout histograms using a Markov chain model. This model contains three parameters, the probability of correctly observing an electron blip when driving the peaks corresponding to the nuclear spin's true state ($p_{\mathrm{corr}}$), the probability of incorrectly observing an electron blip when driving the peaks corresponding to the opposite of the nuclear spin's true state ($p_{\mathrm{err}}$), and the probability of the nuclear spin flipping during the shot ($p_{\mathrm{flip}}$). We find these parameters by fitting the model to experimental readout histograms. The histograms along with the fits and the extracted parameters are shown in Fig.~\ref{fig:supp_nondemo}c-e. The non-zero counts between the peaks correspond to nuclear spin flips during the nuclear spin readout. After obtaining $p_{\mathrm{corr}}$, $p_{\mathrm{err}}$ and $p_{\mathrm{flip}}$, we vary the number of readout shots for each nuclear spin from 1-50, and calculate the individual nuclear spin readout fidelities ($F_1$,$F_2$,$F_3$) from the modelled readout histograms. To obtain the optimal number of shots that maximizes the combined nuclear spin readout fidelity ($F_N$=$F_1 F_2 F_3$), we first note that due to the extremely low probability of nuclear spin 3 flipping during a shot ($p_{\mathrm{flip},3} \approx 0.01$ $\%$), the readout fidelity for nuclear spin 3 saturates after $\sim 19$ shots. We therefore set $N_3=19$. We find that the highest modelled fidelity $F_N=97.90\%$ occurs at ($N_1$,$N_2$,$N_3$)=(19,13,19), and using these values as a guide, we find experimentally that the highest fidelity occurs in a similar parameter regime: ($N_1$,$N_2$,$N_3$)=(24,18,24). We use these values for all measurements that are sensitive to the readout fidelity (QST of the Bell states and of the GHZ state, as well as Grover's algorithm). When measuring all spins at the end of a circuit, the spins are measured in the order (spin 2, spin 3, spin 1), so that spins with higher error are measured first before large errors can accumulate and therefore reduce measurement fidelity. Similarly, when measuring all spins at the start of a circuit (for example to verify that initialisation was successful), spins are measured in the order (spin 1, spin 3, spin 2) to ensure the verification of spins with highest error occurs as close to the start of the circuit as possible, minimising the time for large errors to occur between verification and the circuit.\\

To measure the readout fidelity for a given $N_1$, $N_2$ and $N_3$, we perform an experiment where we read out all three nuclear spins, wait for $\SI{100}{\mu s}$, and then read out all three nuclear spins again (see Fig.~\ref{fig:supp_nondemo}b). This sequence is repeated 100,000 times. We define the readout fidelity as the proportion of repetitions where the first and second readout yield the same nuclear spin configuration in relation to the total number of repetitions. In Fig.~\ref{fig:supp_nondemo}f we show the individual and combined nuclear spin readout fidelities as a function of the relative certainty threshold for ($N_1$,$N_2$,$N_3$)=(24,18,24). The error bars used in Fig.~\ref{fig:supp_nondemo}f are calculated so as to show the range of possible true fidelities which would provide the observed fidelity within $1\sigma$, assuming the measurement results are binomially distributed. Explicitly, they are calculated according to the formula:

\begin{align}
    \sigma_f = \frac{\sqrt{1 + 4 f (1-f) n}}{2(n+1)}
\end{align}

where $\sigma_f$ is the calculated uncertainty, $f$ is the experimentally sampled fidelity, and $n$ is the total number of measurements remaining after certainty thresholding. In Fig.~\ref{fig:supp_nondemo}f, certainty thresholds for individual spins are scaled by factors of ($\frac{1}{2}, 1, \frac{2}{3}$) respectively, postselecting spin 2 more strictly than the others because of its higher error. Throughout this work we use $f_{\mathrm{th}}=0.24$ for spin 2 (and relative scaled values for other nuclei), retaining $\sim 33\%$ of readout measurements with readout fidelities of 99.46\%, 99.42\%, and 99.57\%. 

\clearpage

\section{Nuclear spin initialization}~\label{sec:est}

To initialize the nuclear spins into the desired state (e.g. $\ket{\D\D\D}$), we use a process called electron state transfer (EST) consisting of a sequence of ESR and NMR pulses, previously demonstrated in nitrogen vacancies in diamond \cite{Waldherr_2014}. EST is performed at the beginning of each circuit and therefore repeated for every circuit repetition.\\

EST initializes the nuclear spins sequentially. In order to initialize the first nuclear spin, we use the following sequence: starting from an unknown nuclear spin state, we first initialize the electron spin into the down state. Then we apply the four adiabatic ESR pulses that correspond to flipping the electron spin conditional on the first nuclear spin being in the unwanted state. This is followed by an NMR $\pi$-rotation conditional on the electron spin being in the up state. The combination of electron spin down initialization, the four ESR pulses and the NMR pulse flips the first nuclear spin if it is in the unwanted state and leaves it untouched otherwise (since then the electron spin is not flipped to the up state by the ESR pulses). We repeat this sequence for the other two nuclear spins, leading to a fully initialized nuclear spin register. To verify that the nuclear spins are indeed in the correct state, we perform a non-demolition readout of the nuclear spins.\\

Since a verification readout is performed to ensure that the nuclear spins are initialised correctly, after postselecting on the verification readout the nuclear spin initialisation fidelities are the same as the nuclear spin readout fidelities. As discussed in Supplementary section \ref{sec:supp_readout}, this means that postselected nuclear spin initialisation fidelities are above $99$\% for all nuclear spins. 

\section{Dephasing time of the qubits~\label{sec:dephasing}}

\begin{figure*}[h!]
  \includegraphics[width=1.0\textwidth]{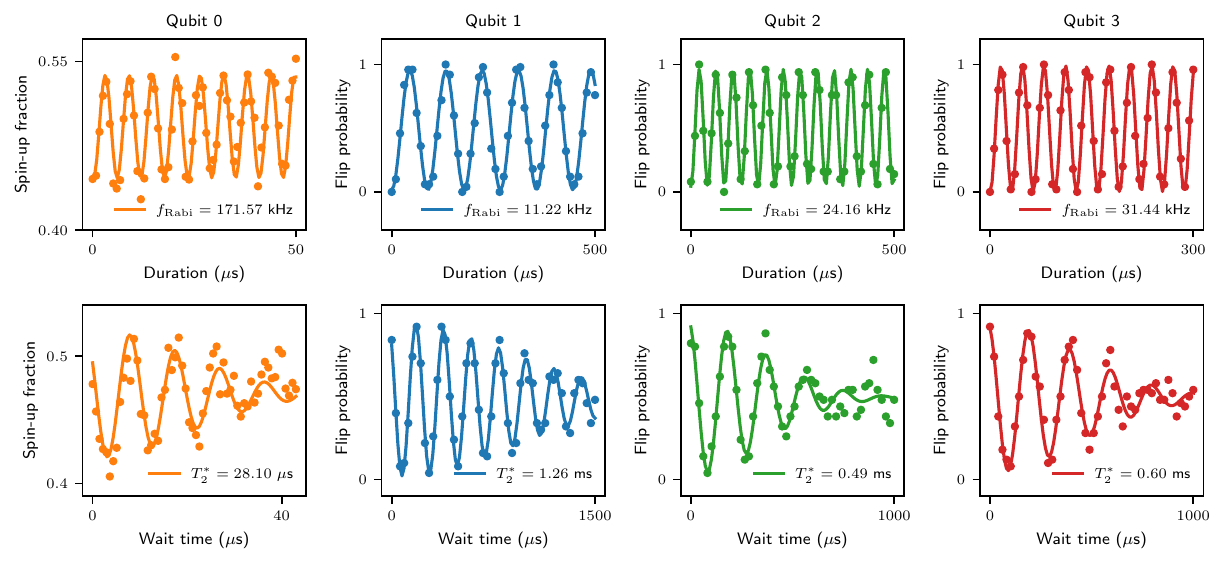}
  \caption{\textbf{Rabi oscillations (top row) and Ramsey measurement (bottom row) for each qubit.} The Rabi frequencies and $T_2^*$ times are indicated for each qubit. For the electron spin qubit, the nuclear spins were initialized into the $\ket{\D\D\D}$ state; for the nuclear spin qubits, the electron spin was initialized into the $\ket{\d}$ state.}
  \label{fig:supp_rabis_ramseys}
\end{figure*}

The dephasing times of the four qubits that we quote in the main text are extracted from Ramsey experiments. For the electron spin qubit, we first initialize the nuclear spins into the $\ket{\D\D\D}$ state using EST, then we initialize the electron spin into the $\ket{\d}$ state, followed by an $R_x(\pi/2)$ rotation, an identity gate with varied wait time, another $R_x(\pi/2)$ rotation, and finally readout of the electron spin (all gates are applied conditional on the nuclear spins being in the $\ket{\D\D\D}$ state). For the nuclear spin qubits, we start by performing a non-demolition readout of the nuclear spins to determine the initial nuclear spin state, followed by the initialization of the electron spin into the $\ket{\d}$ state. We then perform the Ramsey sequence with NMR gates conditional on the electron spin being in the $\ket{\d}$ state, and a final non-demolition readout.\\

We fit the Ramsey measurements according to

\begin{align}
    A\cdot\mathrm{sin}\left(\omega t+\phi \right)\mathrm{exp}\left(-\left(t/T_2^*\right)^2\right)+B,
\end{align}

\noindent where $t$ is the wait time, and $A, \omega, \phi$ and $B$ are fitting parameters. Rabi measurements are fit according to

\begin{align}
    A\cdot\mathrm{sin}\left(\omega t+\phi \right)+B,
\end{align}

\noindent the same form as the Ramsey measurements, but without the decay. Figure~\ref{fig:supp_rabis_ramseys} shows the Ramsey measurements and the fits for the four qubits (bottom row) along with Rabi oscillations for each qubit (top row).\\

We also measured the Rabi frequency and the dephasing time for the electron spin with the nuclear spins initialized into all other configurations, and the results are summarized in Table~\ref{tab:rabis_ramseys}. Note that due to frequency-dependent attenuation in the cables within the dilution refrigerator, the drive amplitudes for each ESR peak have been adjusted to have all ESR rabi frequencies in the proximity of 170 kHz, and that these settings are used throughout the rest of this work. In addition, the Rabi dephasing time was measured for all qubits, as summarised in Table \ref{tab:rabis_ramseys}.\\

\begin{table}[h!]
\centering
\begin{tabular}{|c|c|c|c|} 
 \hline
 \makecell{\Gape[5pt]{Qubit}} & \makecell{\Gape[5pt]{$f_{\mathrm{Rabi}}$ (kHz)}} & \makecell{\Gape[5pt]{$T_2^*$ ($\mu$s)}} & \makecell{\Gape[5pt]{$T_2^{\mathrm{Rabi}}$ (ms)}}\\ 
 \hline
 e ($\ket{\D\D\D}$) & 171.57 & 28.10 & 0.191\\
 \hline
 e ($\ket{\D\D\U}$) & 170.67 & 31.43 & -\\
 \hline
 e ($\ket{\D\U\D}$) & 172.27 & 33.60 & -\\ 
 \hline
 e ($\ket{\D\U\U}$) & 172.01 & 30.79 & -\\ 
 \hline
 e ($\ket{\U\D\D}$) & 168.63 & 26.71 & -\\ 
 \hline
 e ($\ket{\U\D\U}$) & 171.04 & 38.26 & -\\ 
 \hline
 e ($\ket{\U\U\D}$) & 170.64 & 37.75 & -\\ 
 \hline
 e ($\ket{\U\U\U}$) & 171.29 & 26.73 & -\\ 
 \hline
 n1 & 11.22 & 1260 & 71.75\\ 
 \hline
 n2 & 24.16 & 490 & 3.50\\ 
 \hline
 n3 & 31.44 & 600 & 2.21\\ 
 \hline
\end{tabular}
\caption{\textbf{Rabi frequency, dephasing time and Rabi dephasing time for the electron and nuclear spin qubits.} For the electron spin, Rabi and dephasing times are measured with the nuclear spins initialized into the different configurations as depicted. Note that the Rabi dephasing time for the electron spin was only measured for the nuclear state $\ket{\Downarrow\Downarrow\Downarrow}$. The $T_2^*$ times for the nuclear spin qubits are given to the nearest 10 $\mu$s.}
\label{tab:rabis_ramseys}
\end{table}

\section{Single-qubit Randomized Benchmarking}

\begin{figure*}[h!]
  \includegraphics[width=1.0\textwidth]{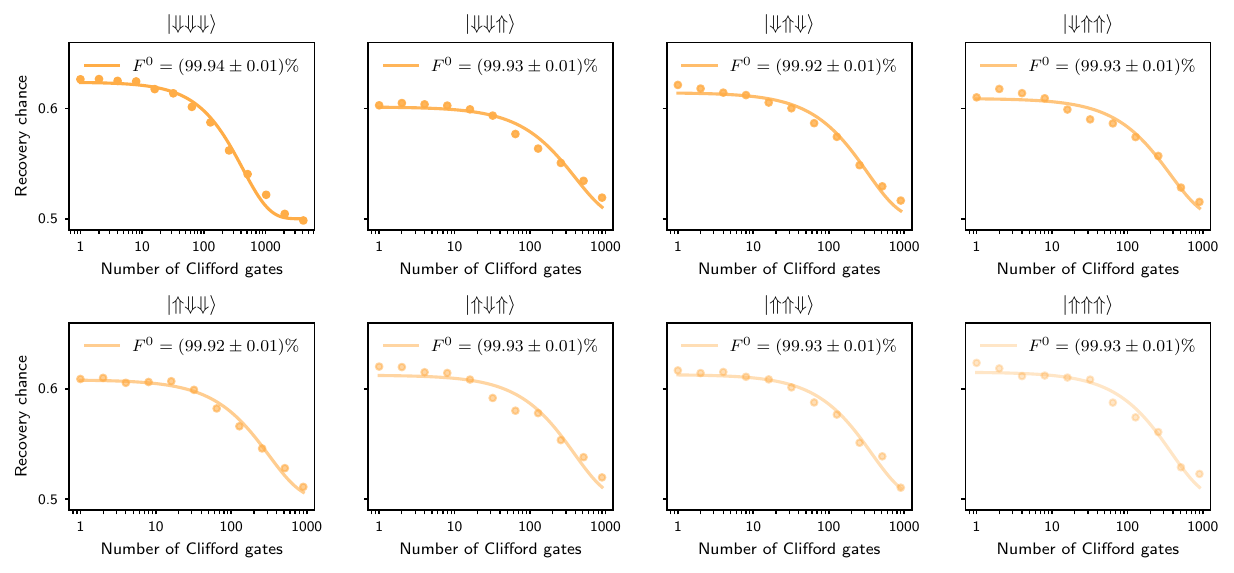}
  \caption{\textbf{Electron spin randomized benchmarking with the nuclear spins initialized into all possible configurations.} The errors for the fidelities are obtained from the fits.}
  \label{fig:supp_ESR_RB}
\end{figure*}

To perform single-qubit randomized benchmarking for the electron spin qubit or the nuclear spin qubits, we first initialize all spins. Afterwards we apply a specific number ($N$) of randomly chosen Clifford gates (each Clifford gate consists of 1.875 physical gates on average), followed by a recovery gate to spin up in the first circuit repetition and spin down in the second repetition, and a readout of the electron or nuclear spin. We then repeat this sequence for varying $N$. Finally, the whole experiment is repeated a number of times (20 for the electron spin and 15 for the nuclear spins), where in every repetition a new randomly chosen set of Clifford gates is applied for every $N$.\\

From this experiment we obtain two decay curves (after averaging over all random variations per $N$), one with recovery to spin up ($P^{\mathrm{u}}$) and one with recovery to spin down ($P^{\mathrm{d}}$). We combine the two curves into a single one according to

\begin{align}
    P = (P^{\mathrm{u}}+(1-P^{\mathrm{d}}))/2,
\end{align}

\noindent which we fit with $af^N+0.5$, and obtain the Clifford gate fidelity from $F_{\mathrm{C}}^i=1-(1-f)/2$, where $i=0,1,2,3$ labels the qubit~\cite{Muhonen_2015}. We then calculate the physical gate fidelity from $F^i=1-(1-F_{\mathrm{C}}^i)/1.875$.\\

In Fig.\ 1d of the main text we show RB for the electron spin with the nuclear spins initialized into the $\ket{\D\D\D}$ state. We also performed RB with the nuclear spins initialized into all other states (using EST as described in Supplementary section~\ref{sec:est}), where we find Clifford gate fidelities above 99\% for all nuclear spin configurations, as shown in Fig.~\ref{fig:supp_ESR_RB}.\\

\section{Quantum State Tomography}

To perform quantum state tomography (QST) on two or three nuclear spin qubits, we measure the qubits in all possible two- or three-qubit Pauli-bases respectively, which we achieve by applying single-qubit NMR rotations prior to the nuclear non-demolition readout. To measure a nuclear spin in the $x$-basis we perform a nuclear $R_{-y}(\pi/2)$ rotation conditional on the electron spin being in the $\ket{\d}$ state, to measure in the $y$-basis we apply a $R_{x}(\pi/2)$ rotation conditional on the electron spin being in the $\ket{\d}$ state, and to measure in the $z$-basis we apply no rotation prior to the non-demolition readout. We apply the rotations in the order (qubit 3, qubit 2, qubit 1), to minimise qubit idle/dephasing time by performing slower rotations first, and perform non-demolition readout of the nuclear spins in the order (qubit 2, qubit 3, qubit 1) so that nuclear spins with higher error are measured first before significant errors can accumulate.\\

To obtain the density matrix from the tomography counts, we perform a constrained Gaussian linear least-squares fit to the count data. The errorbars are obtained from Monte Carlo bootstrap re-sampling and represent $1\sigma$ from the mean~\cite{Watson_2018,Huang_2019}.\\ 

\section{Density matrices for all Bell states}

\begin{figure*}[h!]
  \includegraphics[width=1.0\textwidth]{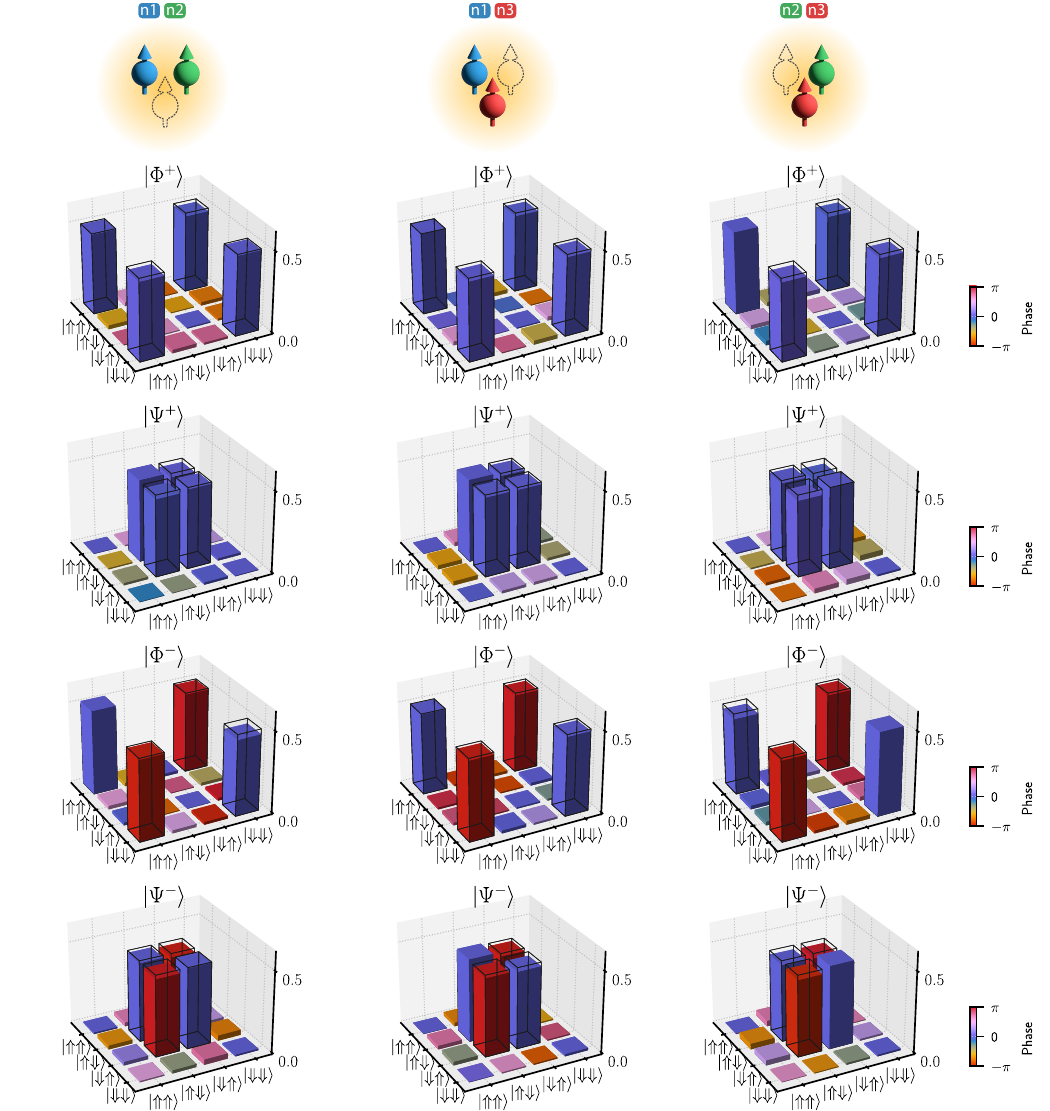}
  \caption{\textbf{Density matrices for all Bell states, for all pairs of nuclear spins.} The left column corresponds to nuclear spin qubit 1 and 2, the middle column to qubit 1 and 3, and the right column to nuclear spin qubit 2 and 3.}
  \label{fig:bell_states}
\end{figure*}

In Fig.\ 2b-d of the main text we show the density matrix for one of the Bell states ($\Phi^+$), for all pairs of nuclear spins. Figure~\ref{fig:bell_states} shows the density matrices for all Bell states ($\Phi^+$, $\Psi^+$, $\Phi^-$, $\Psi^-$), for all pairs of nuclear spins.

\section{Two-qubit Randomized Benchmarking}

To perform two-qubit RB we first initialize the nuclear spins into the $\ket{\D\D\D}$ state, followed by an initialization of the electron spin into the $\ket{\d}$ state. Then we apply a fixed number ($N$) of Clifford gates (each Clifford gate consists on average of $\sim 5.03$ single-qubit $\pi/2$ rotation gates and 1.5 two-qubit CZ gates) and a recovery gate so that the total unitary is identity, returning the two nuclear spins to $\ket{\D}$. The decomposition of Cliffords into native operations is optimised to firstly minimise the number of CZ gates applied, and secondarily to avoid applying single-qubit gates to only one qubit for an extended period where possible (to avoid the other qubit dephasing). The sequence of $N$ Clifford gates is followed by non-demolition readout of the nuclear spins. This sequence is repeated for varying $N$. Afterwards we repeat the experiment with new randomly chosen Clifford gates for every $N$. After 5 of these repetitions we perform 5 repetitions where we interleave a two-qubit CZ gate between every Clifford gate to perform interleaved two-qubit RB. Finally, we repeat the sequence of 5 non-interleaved and 5 interleaved repetitions 9 more times to arrive at a total of 50 non-interleaved and 50 interleaved final runs for each pair of nuclear spins.\\

From this measurement we obtain two RB decay curves (after averaging over all random variations per $N$), a non-interleaved (reference) curve and an interleaved curve. We fit both curves with $af^N+b$ and calculate the corresponding Clifford fidelity from $F_{\mathrm{ref/int}}^{ij}=1-3(1-f_{\mathrm{ref/int}})/4$, where $i,j \in [1,2,3]$ label the used nuclear spins. From the decay rate ratio ($d=f_{\mathrm{int}}/f_{\mathrm{ref}}$) we obtain the interleaved gate fidelity via $F_{\mathrm{CZ}}^{ij}=1-3(1-d)/4$~\cite{Xue_2019}.\\

\section{Nuclear-Nuclear Two-Qubit Gate Errors}

In this work, the multi-qubit gate between nuclear spins is achieved by performing a $2\pi$ rotation of the electron, conditional on a particular nuclear spin configuration, as discussed in the main text. The fidelity of this multi-qubit gate was characterised in Fig.\ 2 of the main text between each pair of nuclei, giving fidelities of $99.65 \pm 0.35$ \%, $99.49 \pm 0.39$ \% and  $99.32 \pm 0.22$ \%, giving an average value of $99.49 \pm 0.20$ \%. Here, we explore how quasistatic variation of the electron energy splitting affects the performance of the $2\pi$ rotation, and show that the corresponding predicted error agrees with the observed experimental value. For ease of calculation, we work in angular frequency; ie. setting $\hbar = 1$.\\

When the electron is present, and in the absence of control pulses, the Hamiltonian of the multi-nuclear register is given by:

\begin{align}
	H &= \left(\gamma_e S_z + \sum_{i=1}^3 \gamma_n I_{i,z}\right) B_0 + \sum_{i=1}^3 A_i \mathbf{S} \cdot \mathbf{I_i} ~\label{eq:supp_original_hamiltonian}
\end{align}

\noindent where $\gamma_e$ ($\gamma_n$) is the electron (nuclear) gyromagnetic ratio, and $A_i$ is the contact hyperfine coupling interaction strength for nuclear spin $i \in \{1, 2, 3\}$. To model the evolution of all qubits in this system, we work in the interaction picture. That is, we choose the unitary

\begin{align}
	U(t) &= \textrm{exp} \left(i H t\right)
\end{align}

\noindent and use it to transform the Hamiltonian and state as follows:

\begin{align}
	\ket{\psi(t)} &\to U(t) \ket{\psi(t)}\\
	H &\to i \frac{\partial U(t)}{\partial t} U^\dag(t) + U(t) H U^\dag(t) = 0
\end{align}

\noindent Working in the interaction picture ensures that when the qubits are idle, they will not pick up phase as the Hamiltonian is 0. We define the phase of our qubits in the interaction frame, so that any phase errors accumulated will be solely caused by errors in the magnetic control pulses.\\

To model the action of a control pulse, it is useful to temporarily tranform the system to a rotating frame which rotates at the frequency of the control pulse, making sure to return back to the original interaction frame at the end of the control pulse to establish the final phase of the qubits. As we are interested in the application of ESR control in this section, it is also valid to work in the secular approximation as $\gamma_e B_0 \sim 39$ GHz is much larger than other energy scales of the Hamiltonian (that is, we assume only the $z$-terms contribute in the Hamiltonian in Eq. \ref{eq:supp_original_hamiltonian}). In addition, we apply the rotating wave approximation, as we are in the limit of weak driving. Under these approximations, for a two-level system with energy splitting $E$ and a control pulse of frequency $E + \delta$ and drive strength $\omega$, the Hamiltonian in a frame rotating at $E + \delta$ is given by:

\begin{align}
	H_d &= - \frac{\delta}{2} \sigma_z + \frac{\omega}{2} \sigma_x
\end{align}

\noindent After applying this drive Hamiltonian for some time $t$, the unitary applied is given by:

\begin{align}
	U_d &= \textrm{exp} \left(-i H t\right)\\
	&= \sigma_I \cos\left(\sqrt{\delta^2 + \omega^2} t/2\right) + i \frac{\delta \sigma_z - \omega \sigma_x}{\sqrt{\delta^2 + \omega^2}} \sin \left(\sqrt{\delta^2 + \omega^2} t/2\right)
\end{align}

\noindent If the control pulse is far off-resonance, ie. $\delta \gg \omega$, we have

\begin{align}
	U_d'(t) &\approx \sigma_I \cos\left(\delta t/2\right) + i \sigma_z \sin \left(\delta t/2\right)\\
	U_d'(t) &\approx \textrm{exp} \left(i \delta \sigma_z t/2\right) ~\label{eq:supp_far_offres_drive_ham}
\end{align}

Now, consider the application of a $2\pi$ rotation of the electron in the 4-qubit register considered in this work, conditional on the nuclear state $\ket{\Downarrow\Downarrow\Downarrow}$, with the electron starting $\ket{\downarrow}$. If the terms in Eq. \ref{eq:supp_original_hamiltonian} were known exactly, then such a control pulse (in the secular approximation) would be applied at a frequency of $\gamma_e B_0 - \left(\sum_i A_i\right)/2$, which is the energy splitting between $\ket{\downarrow\Downarrow\Downarrow\Downarrow}$ and $\ket{\uparrow\Downarrow\Downarrow\Downarrow}$ in the secular approximation. However, due to imprecise knowledge of the Hamiltonian, the applied frequency will actually be $\gamma_e B_0 - \left(\sum_i A_i\right)/2 + \delta$ for some small detuning $\delta$ (representing the difference between the experimentally calibrated energy splitting and the true energy splitting). To transform from the interaction picture to a frame rotating at this drive frequency, for the subspace $\left\{\ket{\downarrow\Downarrow\Downarrow\Downarrow}, \ket{\uparrow\Downarrow\Downarrow\Downarrow} \right\}$, one applies the unitary:

\begin{align}
	U_{\textrm{trans}} (t) &= \textrm{exp} \left(i \delta S_z t\right)
\end{align}

\noindent to account for the frequency difference $\delta$ between the interaction picture and our drive frequency. We assume that the pulse starts at $t=0$, and finishes after the electron has returned to the $\ket{\downarrow}$ state at time $t = \tau \equiv 2\pi/\omega \approx 2\pi/\sqrt{\delta^2 + \omega^2}$. Then the full unitary that is performed is given by:

\begin{align}
	U_{2\pi} &= U_{\textrm{trans}}^\dag \left(\tau\right) U_d \left(\tau\right) U_{\textrm{trans}} (0)\\
	&= U_{\textrm{trans}}^\dag \left(\tau\right) \cdot (-I) \cdot (I)\\
	U_{2\pi} &= -\begin{bmatrix}
		e^{-i \delta \tau / 2} & 0\\
		0 & e^{i \delta \tau / 2}
	\end{bmatrix}
\end{align}

\noindent which is the unitary applied to the subspace $\left\{\ket{\downarrow\Downarrow\Downarrow\Downarrow}, \ket{\uparrow\Downarrow\Downarrow\Downarrow} \right\}$. For all other pairs of states, the drive will be far from the energy splitting, so we instead use the drive Hamiltonian when far off-resonance (given in Eq. \ref{eq:supp_far_offres_drive_ham}), which gives:

\begin{align}
	U_{2\pi}' &= U_{\textrm{trans}}^\dag \left(\tau\right) U_d' \left(\tau\right) U_{\textrm{trans}} (0)\\
	&\approx \textrm{exp} \left(-i \delta S_z \tau\right) \cdot \textrm{exp} \left(i \delta S_z \tau\right) \cdot (I)\\
	U_{2\pi}' &\approx I
\end{align}

\noindent Hence, for all pairs of states \textit{except} the subspace $\left\{\ket{\downarrow\Downarrow\Downarrow\Downarrow}, \ket{\uparrow\Downarrow\Downarrow\Downarrow} \right\}$, the identity unitary is performed. Therefore, if we project into the electron-$\ket{\downarrow}$ subspace (limiting ourselves to a pair of nuclear spins, as we are concerned with the two-qubit gate fidelity), the unitary is given by:

\begin{align}
	U_{2\pi} &= \begin{bmatrix}
	-e^{-i \delta \tau / 2}&&&\\
	&1&&\\
	&&1&\\
	&&&1
	\end{bmatrix} ~\label{eq:supp_2pi_unitary}
\end{align}

\noindent The fidelity of this operation is given by~\cite{Raginsky_2001}

\begin{align}
	F &= \frac{1}{d^2} \left|\textrm{tr}\left(U_{2\pi}^\dag V_{2\pi}\right) \right|^2
\end{align}

\noindent where $d$ is the dimension of the Hilbert space, and $V_{2\pi}$ is the ideal intended unitary, ie. $U_{2\pi} |_{\delta=0}$. Substituting in Eq. \ref{eq:supp_2pi_unitary}, we get:

\begin{align}
	F &= \frac{1}{16} \left|3 + e^{i \delta \tau / 2} \right|^2\\
	&= \frac{\left[3 + \cos (\delta \tau/2)\right]^2 + \sin^2 (\delta \tau/2)}{16}\\
	&\approx \frac{\left[4 - \frac{\delta^2 \tau^2}{8}\right]^2 + (\delta \tau/2)^2}{16} + O(\delta^4 \tau^4)\\
	&\approx \frac{16 - \delta^2 \tau^2 + (\delta \tau/2)^2}{16} + O(\delta^4 \tau^4)\\
	F &\approx 1 - \frac{3}{4} \left(\frac{\delta \tau}{4}\right)^2 + O(\delta^4 \tau^4)
\end{align}

\noindent Hence, the average two-qubit error is given by

\begin{align}
	\epsilon &= 1 - F\\
	\epsilon &\approx \frac{3}{4} \left(\frac{\delta \tau}{4}\right)^2 + O(\delta^4 \tau^4)
\end{align}

In practice, the detuning $\delta$ will not be fixed, but instead will vary over time due to noise in the energy splitting of the electron. A significant component of this variation is quasistatic, meaning $\delta$ will vary repetition-to-repetition in an amount characterised by $T_2^*$. Specifically, the variance in $\delta$ (measured in angular frequency) caused by $T_2^*$ is given by~\cite{Dial_2013}:

\begin{align}
	\textrm{Var}(\delta) &= \frac{2}{(T_2^*)^2}
\end{align}

\noindent The expected error can be calculated as follows:

\begin{align}
	\textrm{E}(\epsilon) &= \int \epsilon(\delta) p(\delta) d\delta\\
	&\approx \int \frac{3}{4} \left(\frac{\delta \tau}{4}\right)^2 p(\delta) d\delta\\
	\textrm{E}(\epsilon) &\approx \frac{3}{4} \left(\frac{\tau}{4}\right)^2 \textrm{E}(\delta^2)
\end{align}

\noindent where E denotes expectation value. Assuming that the experiment is well-calibrated, meaning that $\textrm{E}(\delta) = 0$, we can utilise the fact that $\textrm{Var}(\delta) = \textrm{E}(x^2) - \left[\textrm{E}(x)\right]^2$ to show that $\textrm{E}(\epsilon)$ is inversely proportional to the square of the dephasing time:

\begin{align}
	\textrm{E}(\epsilon) &\approx \frac{3}{4} \left(\frac{\tau}{4}\right)^2 \frac{2}{(T_2^*)^2}\\
	\textrm{E}(\epsilon) &\approx \frac{3}{32} \left(\frac{\tau}{T_2^*}\right)^2 ~\label{eq:supp_final_error_formula}
\end{align}

\noindent Equation \ref{eq:supp_final_error_formula} gives us a direct method of predicting the fidelity of the CZ operation. Substituting in the average values given in table~\ref{tab:rabis_ramseys} of $f_{\textrm{Rabi}} = 171.0 \pm 1.1$ kHz and $T_2^* = 31.7 \pm 4.3$ $\mu$s, we get a theoretical error (and fidelity) of:

\begin{align}
	\textrm{E}(\epsilon) &= 0.32 \pm 0.09 \%\\
	\textrm{E}(F) &= 99.68 \pm 0.09 \%
\end{align}

\noindent which is in agreement with the experimental average value of $F = 99.49 \pm 0.20$ \%. While within uncertainties, the experimental fidelity could potentially include effects from miscalibration (ie. $E(\delta) \neq 0$), and also effects from dephasing of the nuclear spins during the application of the $2\pi$ ESR pulse. Despite this, Eq. \ref{eq:supp_final_error_formula} provides an upper bound on the qubit fidelity. Equation \ref{eq:supp_final_error_formula} also indicates quantities to optimise in order to increase two-qubit fidelities further, namely by aiming to extend electron $T_2^*$ times and reduce electron gate times $\tau$.\\

\section{Additional data for Grover's algorithm}

\begin{figure*}[h!]
  \includegraphics[width=1.0\textwidth]{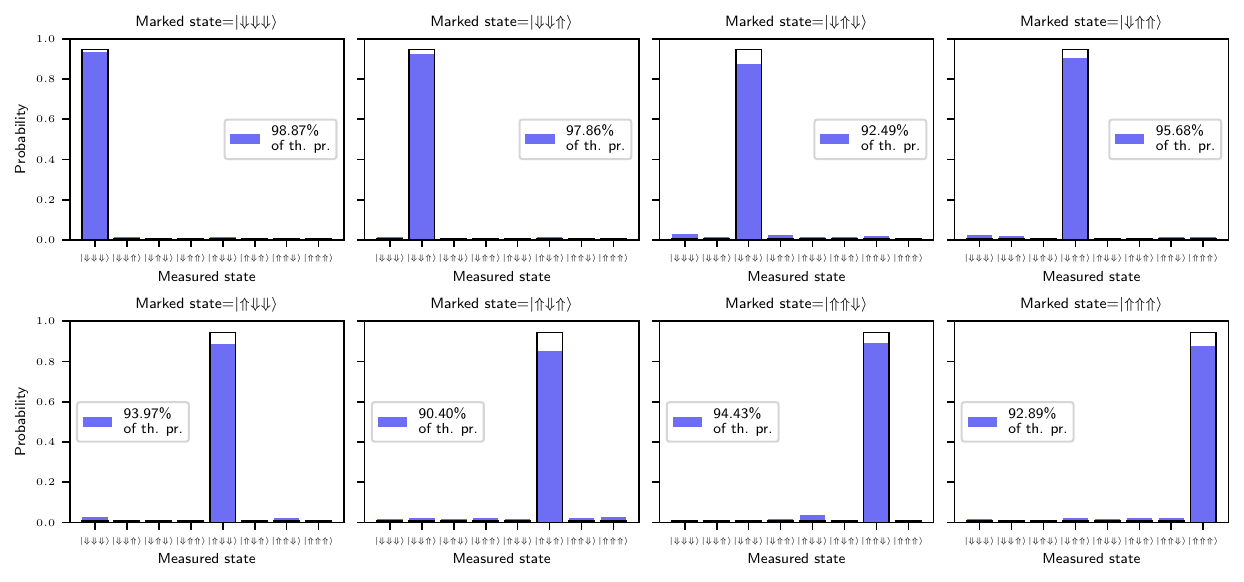}
  \caption{\textbf{Grover's algorithm executed with all possible marked states.} For each measurement we indicate the probability of finding the correct marked state normalized by the theoretical maximum probability of finding the marked state with two Grover iterations (94.53\%).}
  \label{fig:supp_grovers}
\end{figure*}

In Fig.\ 4b of the main text we show Grover's algorithm executed with $\ket{\D\D\D}$ as the marked state. We also ran this algorithm using all other possible states as the marked state, and the results are shown in Fig.~\ref{fig:supp_grovers}. We achieve an average probability of ($89.40\pm2.49)\%$ of measuring the marked state, which corresponds to ($94.57\pm2.63)\%$ of the theoretical maximum value (94.53\%).\\

\section{Error budget for Grover's Algorithm~\label{sec:supp_grover_error_budget}}

In this work, we perform Grover's search algorithm over all potential marked states $m$, as shown in Fig.\ 4 of the main text and supplementary information Fig.~\ref{fig:supp_grovers}. On average, the success probability of our algorithm is ($94.57 \pm 2.63$)\% of the ideal success probability. In Table \ref{tab:supp_grovers_errorbudget}, we list the operations needed to implement the algorithm as per Fig.\ 4a of the main text, along with the number of times that operation is used. In addition, we include the error of each unique operation, as characterised in the main text (for single- and multi-qubit operations) and in supplementary information section~\ref{sec:supp_readout} (for SPAM errors). By multiplying these fidelities together, according to multiplicities given by the number of applications of each operation, we obtain an estimate of the average fidelity of the algorithm as $95.89$ \% (also shown in Table \ref{tab:supp_grovers_errorbudget} as "Total (no idle errors)"), matching closely with the average experimental performance of the algorithm.

\begin{center} 
\begin{table}[htbp!]
\centering
\begin{tabular}{| c | c | c |}
	\hline
	\makecell{\Gape[7pt]{Operation}} & \makecell{\Gape[7pt]{Num. Occurences}} & \makecell{\Gape[7pt]{Fidelity}}\\\hline
	SPAM (n1) & 1 & 99.46\%\\\hline
	SPAM (n2) & 1 & 99.42\%\\\hline
	SPAM (n3) & 1 & 99.57\%\\\hline
	Single-qubit gate (n1) & 5 & 99.98\%\\\hline
	Single-qubit gate (n2) & 5 & 99.95\%\\\hline
	Single-qubit gate (n3) & 5 & 99.95\%\\\hline
	Multi-qubit gate & 4 & 99.49\%\\\hline
	\textbf{Total (w/o idle errors)} & - & \textbf{95.89\%}\\\hline
	n1 idle during $\pi/2$ rotations on n2 \& n3 & 4 & 99.98\%\\\hline
	n2 idle during $\pi/2$ rotations on n1 \& n3 & 4 & 99.62\%\\\hline
	n3 idle during $\pi/2$ rotations on n1 \& n2 & 4 & 99.70\%\\\hline
	\textbf{Total (w/ idle errors)} & - & \textbf{93.23\%}\\\hline
\end{tabular}

\caption{\textbf{Error budget for the implementation of Grover's search algorithm performed in this work.} The fidelities in the rightmost column for gate errors are as characterised in the main text of this work and supplementary section~\ref{sec:supp_readout}. The "Total (w/o idle errors)" row is found by multiplying together the above fidelities, according to multiplicities given by the number of occurrences of each operation. The idle fidelities are calculated as per the formula $f \approx \textrm{exp}\left(-\left(\tau/T_2^*\right)^2\right)$, outlined in this supplementary section. The total fidelity with idle errors is the product of these idling fidelities (according to their multiplicities) and the total fidelity without idle errors.}
\label{tab:supp_grovers_errorbudget}
\end{table}
\end{center}

The error budget given in Table \ref{tab:supp_grovers_errorbudget} without idling errors slightly overestimates the observed fidelity of 94.57 \% measured in this work, likely because the single-qubit gate fidelities do not include idling errors incurred on the other nuclear qubits during the single-qubit operation. From supplementary section~\ref{sec:dephasing} we know that the nuclear spins experience noticeable dephasing of $T_2^* = $ 1.26 ms, 0.49 ms and 0.60 ms for nuclear spins n1, n2 and n3 respectively. During control of n1, for example, n2 and n3 will experience $T_2^*$ dephasing according to the duration of the gate on n1. Note that such dephasing will not impact fidelity if n2 or n3 are in a z-basis state (such as $\ket{\Downarrow}$, as they are at the start of the algorithm); and similarly, dephasing will not affect fidelity at the end of the circuit immediately before the final measurement. For the remaining idle times, we can approximate the phase error from the formula $f \approx \textrm{exp}\left(-\left(\tau/T_2^*\right)^2\right)$, where $\tau$ is the total time spent idling. Using this formula and the rabi frequencies given in supplementary section~\ref{sec:dephasing}, we append additional errors to Table \ref{tab:supp_grovers_errorbudget}. The overall total fidelity (including these idling errors) of 93.24 \% is again close to the observed average of 94.57 \%, but now underestimates the fidelity. Resolving the discrepancy between these two values is beyond the scope of this work, but suggests that the implementation of Grover's algorithm performed in this work is to some extent resilient to $T_2^*$ errors, potentially due to the regularity of $\pi/2$ rotations on all 3 qubits creating partial refocusing of the nuclear spins.

\begin{figure*}[h!]
  \includegraphics[width=1.0\textwidth]{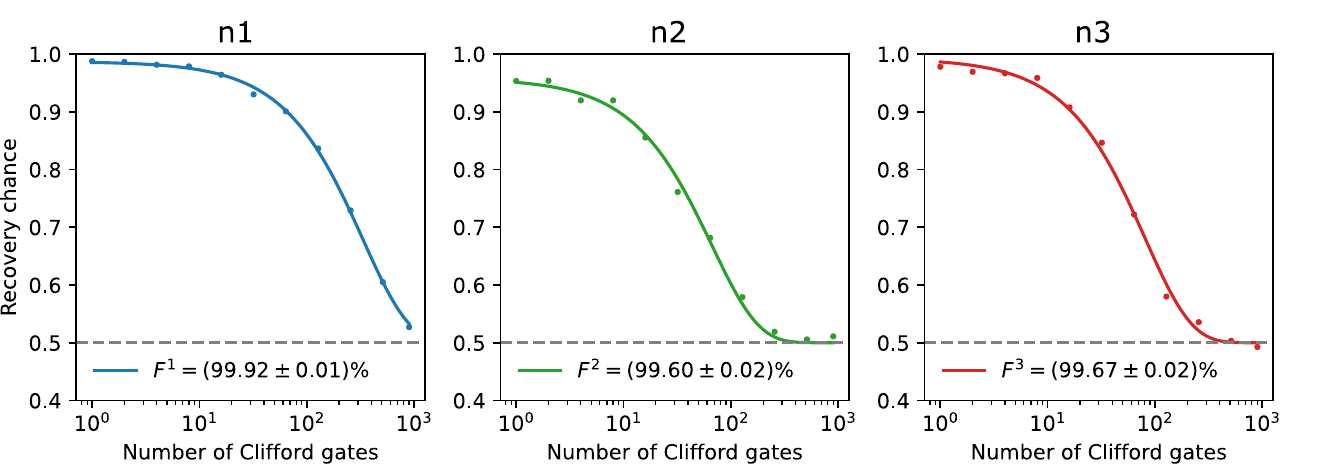}
  \caption{\textbf{Modified single-qubit randomised benchmarking on the nuclear spins.} Here, randomised benchmarking is performed on all 3 nuclear spins in the same circuit, by performing physical gates on each of the nuclear spins sequentially (see text in this section for details). The reduced physical gate fidelities measured here (compared to randomised benchmarking performed individually on each nuclear spin, as shown in Fig.\ 1d of the main text) are modelled well by considering the extra idling time, and hence $T_2^*$ dephasing, as calculated in Table \ref{tab:supp_grovers_errorbudget}. The errors for the fidelities are obtained from the fits.}
  \label{fig:supp_sequential_1Q_RB}
\end{figure*}

To further characterise the errors incurred during idling, we perform a modified single-qubit randomised benchmarking experiment. For each randomised benchmarking variation, we generate 3 independent randomised benchmarking sequences, one for each of the nuclear spin qubits. These sequences are then converted to a sequence of physical gates, labelled $\{g_1^{n1}, g_2^{n1}, g_3^{n1}, \dots, g_{m_1}^{n1}\}$ for nuclear spin 1, $\{g_1^{n2}, g_2^{n2}, \dots, g_{m_2}^{n2}\}$ for nuclear spin 2, and likewise for nuclear spin 3. In the modified single-qubit randomised benchmarking circuit, we apply these gates sequentially over the three nuclear spins, ie in the order $g_1^{n1}, g_1^{n2}, g_1^{n3}, g_2^{n1}, g_2^{n2}, \dots$ (sequentially doing physical gates to each qubit in turn). If all physical gates for one nuclear spin have been exhausted (possible if eg. $m_1 \neq m_2$), that nuclear spin will be skipped until all nuclear spins have completed their respective full sequences of gates. The results of this experiment are shown in Fig. \ref{fig:supp_sequential_1Q_RB}, showing fidelities of ($99.92 \pm 0.01$)\%, ($99.60 \pm 0.02$)\%, and ($99.67 \pm 0.02$)\% for nuclear spins 1, 2 and 3 respectively. This is in close agreement with the idle fidelities shown in Table. \ref{tab:supp_grovers_errorbudget}, again supporting the idea that nuclear fidelities can be described accurately with (standard) randomised benchmarking fidelities, along with $T_2^*$ dephasing effects.

\section{State-of-the-art implementations of Grover's algorithm}

Table \ref{Grover_implementations} summarizes the state-of-the-art experimental implementations of Grover's algorithm reported to-date in the literature across multiple qubit platforms. Grover's algorithm has been demonstrated for up to 5 qubits. However, demonstrations on 4- and 5-qubit systems are limited in success probability ($<50\%$) due to the long circuit lengths required. Our implementation with the marked $\ket{\D\D\D}$ state resulted in a success probability of 93.46\% corresponding to 98.87\% of the theoretical maximum, the highest reported to-date. \\\\\\

\begin{center} 
\begin{table}[htbp!]
\begin{tabular}{| c | c | c | c | c | c | c | c |}
	\hline
	\makecell{\Gape[7pt]{Reference}} & \makecell{\Gape[7pt]{Year}} & \makecell{\Gape[7pt]{Platform}} & \makecell{\Gape[7pt]{Qubits}} & \makecell{\Gape[7pt]{Iterations}} & \makecell{\Gape[7pt]{Success prob.}} & \makecell{\Gape[7pt]{Theoretical max.}} & \makecell{\Gape[7pt]{Ratio}}\\
	\hline

    \makecell{This paper ($\ket{\D\D\D}$ state)\\This paper (average)} & 2024 & Si:P qubits (SQC)
    & \makecell{3\\3} & \makecell{2\\2} & \makecell{93.46\%\\89.40\%} & \makecell{94.53\%\\94.53\%} & \makecell{0.9887\\0.9457}\\
   \hline

    \multirow{2}{*}{Hlembotskyi $\textit{et al.}$ \cite{hlembotskyi2020efficient}} & \multirow{2}{*}{2020} & \multirow{2}{*}{Ion trap qubits (Honeywell)} 
    & 3 & 1 & 75.2\%  &78.1\% & 0.96\\
    &&& 5 & 1 & 18.7\%  &25.8\% & 0.72\\
   \hline
     
    \multirow{3}{*}{Mandviwalla $\textit{et al.}$ \cite{mandviwalla2018implementing}} & \multirow{ 3}{*}{2018} & \multirow{ 3}{*}{Superconducting qubits (IBM)} 
    & 2 & 1 & 80.9\%  &100\% & 0.81\\
    &&& 3 & 2 & 59.7\%  &94.5\% & 0.63\\
    &&& 4 & 3& 6.6\%  & 96.1\% & 0.07\\
   \hline

    Zhang $\textit{et al.}$ \cite{zhang2022quantum} & 2022 & Ion trap qubits (Honeywell) & 5 & 2& 49\%   &60.2\%& 0.81\\  
     \hline

    \multirow{5}{*}{Zhang $\textit{et al.}$ \cite{zhang2021implementation}} & \multirow{5}{*}{2021} & \multirow{5}{*}{Superconducting qubits (IBM)} 
    & 3 & 1 &  55.9\% &78.1\%& 0.72\\
    &&& 3 & 2 & 63.8\% &94.5\%& 0.68\\
    &&& 4 & 1 & 18.1\%&47.3\%& 0.38\\
    &&& 4 & 2 & 19.5\%&90.8\%& 0.21\\
    &&& 5 & 1& 2.6\%   &25.8 \%& 0.10\\
   \hline

	Adedoyin $\textit{et al.}$ \cite{adedoyin2018quantum} & 2018 & Superconducting qubits (IBM) & 2 & 1& 65\%   &100 \%& 0.65\\
   \hline

   Friggatt $\textit{et al.}$ \cite{figgatt2017complete} & 2017 & Ion trap qubits & 3 & 1& 38.9\%   &78.1\%& 0.49\\   
    \hline
    
	Gwinner $\textit{et al.}$ \cite{gwinner2020benchmarking} & 2020 & Superconducting qubits (IBM) & 4 & 1 & 21.0\%   &47.3 \%& 0.44\\
   \hline
     
	Stromberg $\textit{et al.}$ \cite{stromberg2018} & 2018 & Superconducting qubits (IBM) & 4 & 1& 6.6\%   &47.3 \%& 0.14\\
   \hline
   
	Watson $\textit{et al.}$ \cite{Watson_2018} & 2018 & Si/SiGe spin qubits & 2 &  1 & - & 100 \%& -\\
	\hline

\end{tabular}

\caption{\textbf{State-of-the-art experimental implementations of Grover's algorithm in various qubit platforms.} The last column represents the ratio between the measured success probability and the theoretical maximum.}
\label{Grover_implementations}
\end{table}
\end{center}

\clearpage

{\bf References: }

\bibliography{references}
\bibliographystyle{apsrev4-1}

\end{document}